\documentclass[preprint,preprintnumbers,amsmath,amssymb]{revtex4}
\usepackage{amsfonts}
\usepackage{amssymb}
\usepackage{latexsym}
\usepackage{graphicx}% Include figure files
\newcommand{\rar}{\rightarrow}

\begin{document}

\title{An accurate few-parameter ground state wave function for the Lithium
 atom}

\author{Nicolais L.~Guevara}
\email{guevara@qtp.ufl.edu}
\affiliation{Quantum Theory Project, Departments of Physics and Chemistry\\
                University of Florida, Gainesville, FL 32611-8435, USA}

\author{Frank~E.~Harris}
\email{harris@qtp.ufl.edu}
\affiliation{Quantum Theory Project, Departments of Physics and Chemistry\\
                University of Florida, Gainesville, FL 32611-8435, USA}

\author{Alexander V.~Turbiner}
\email{turbiner@nucleares.unam.mx}
\affiliation{Instituto de Ciencias Nucleares, Universidad Nacional
Aut\'onoma de M\'exico, Apartado Postal 70-543, 04510 M\'exico,
D.F., M\'exico}

\bigskip

\begin{abstract}
A simple, seven-parameter trial function is proposed for a description of
the ground state of the Lithium atom. It includes both spin functions.
Inter-electronic distances appear in exponential form as well as in a
pre-exponential factor, and the necessary energy matrix elements are
evaluated by numerical integration in the space of the relative coordinates.
Encouragingly accurate values of the energy and
the cusp parameters as well as for some expectation values are obtained.

\end{abstract}

%\pacs{31.15.Pf,31.10.+z,32.60.+i,97.10.Ld}

\maketitle

\section{Introduction}

It is well-known that the standard quantum chemistry approaches to
calculation of the energies of the low-lying states of  few-electron
atoms are characterized by slow convergence. This convergence problem
was demonstrated in a discouragingly explicit way for
the case of the Helium atom in recent studies by Korobov \cite{korobov}
(using a sort of exponential Hylleraas basis) and by Schwartz \cite{schwartz}
(using various trial functions that included correlated exponentials,
pre-exponential integer and fractional  powers of both nuclear-electron and
 electron-electron distances, as well as logarithmic terms).
An unpleasant drawback of these studies is an
absence of confidence that high accuracy obtained for the energy
guarantees a comparable accuracy in expectation values
%%%%%
% \footnote{It is worth to emphasize that in literature sometimes
% it appears misleading statements
%about accurate wavefunctions without explanations of a meaning of the word
%"accurate": without giving real estimates of accuracy of the wavefunction
%in whatever sense, except for the energy (see e.g. \cite{exact}). In the same
%time there are explicit examples when a straightforward extrapolation to the
%exact values lead to the incorrect results (see e.g. \cite{exact-exp}). }.
%%%%%
\footnote{It is worth emphasizing that in the
literature there sometimes appear misleading statements about accurate
wavefunctions without any explanation of the meaning of the word
``accurate''; as pointed out in \cite{exact} the only accuracy measure usually
identified is the energy.  Moreover, there are explicit examples in
which a straightforward extrapolation of the variational energy (and expectation values) to the exact energy leads to
inaccurate results  (see e.g. discussion in \cite {exact-exp}). }
%%%%%
 This
question is vital when relativistic corrections are studied, particularly
in view of the fact that some of these corrections are defined by the expectation values of singular and/or local quantities { (for a discussion see e.g. \cite{exact})}. Recent advances in experimental techniques
have now led to experimental data whose understanding requires an accurate
knowledge of the relativistic corrections \cite{Eides}. As one of
possible ways to handle this situation, one of the present authors (FEH)
has proposed to look for simple, few-parameter "ultra-compact" trial
functions which guarantee reasonably high overall accuracy.  One way
of characterizing this approach is to describe it as a search for the
most accurate few-parameter trial functions.   This line of endeavor
is illustrated by work on the H$_2$ molecule \cite{H2}, for which a nearly
optimum 14-parameter function was reported, and by work on the He
isoelectronic series \cite{Feh_He}, where optimum wavefunctions of up to
four configurations were generated.

The present contribution deals with a search for an optimum ultra-compact
wavefunction for the ground state of the Li atom.  An important issue
for such a study is how to determine
the overall quality of a trial function.  The viewpoint taken in
the present work is to use as a quality measure the error in the cusp
parameters obtained
from the trial function (residues arising from the Coulomb singularities
of the potential).  Of course, this criterion becomes reasonable only if
the cusp conditions are not artificially fixed to their exact values
by the choice of form for the trial function.  We note that the most
popular methods for atomic computations use Gaussian-type orbitals, and
that while such bases can provide extremely accurate energies, they
usually lead to vanishing cusp parameters and thereby have significant
drawbacks for the description of relativistic and other local effects.

In the work reported here, the trial functions that were examined consist of
exponentials in all the relative coordinates, in some cases multiplied by
pre-exponential factors dependent linearly on the interparticle distances.
The matrix elements that arise have been evaluated numerically by methods
used previously by one of the authors \cite{turbiner:2006}.

\section{Wavefunction and Variational Method}

The nonrelativistic Hamiltonian for the Lithium atom under the
Born-Oppenheimer approximation of zero order, i.e.\ with the Li nucleus
assumed to be of infinite mass, is (in Hartree atomic units)
\begin{equation}
\label{H}
  {\cal H}\ = - \sum_{i=1}^3 \left( \frac{1}{2} {\nabla}_{i}^2\ +
  \  \frac{Z}{r_{i}} \right)
  \ + \ \sum_{{i=1}}^3 \sum_{{j>i}}^3 \ \frac{1}{r_{ij}}\   \, ,
\end{equation}
where ${\nabla}_{i}$ is the 3-vector
of the momentum of the $i$th electron, $Z$ is the nuclear charge
(here Z=3), $r_{i}$ is the distance
between the $i$th electron and the Li nucleus, and $r_{ij}$ are the
interelectron distances.
The kinetic energy part of $\cal H$ is conveniently written in terms of
the distance coordinates $r_i$ and $u_k = r_{ij}, k \neq i \neq j$ \cite{larsson,king:1997},
\begin{eqnarray}
\sum_{i=1}^3 {\nabla_{i}}^2\ &=&  \sum_{i=1}^3
 \left( \frac{\partial^2}{\partial r_i^2}
 +\frac{2}{r_i} \frac{\partial}{\partial r_i}
 + 2 \frac{\partial^2}{\partial u_{i}^2}
 +   \frac{4}{u_i} \frac{\partial}{\partial u_i} \right)    \nonumber\\
&+&  \sum_{P} \left( \frac{r_i^2 + u_k^2 -r_j^2}{u_k r_i}
 \frac{\partial^2}{\partial r_i \partial u_k}  +
\frac{1}{2} \frac{u_i^2 + u_k^2 -u_j^2}{u_i u_k}
  \frac{\partial^2}{\partial u_i \partial u_k}
 \right),
\label{N}
\end{eqnarray}
as long as the wave function has no explicit angular dependence.
The summation $P$ runs over $i,j,k$ which are
the six permutations of $1,2,3$.

The variational method is used to study the ground state of the Lithium atom.
Physical relevance arguments are followed to choose the trial function
(see, e.g.\ Turbiner \cite{turbinervar}). { In particular, we construct wavefunctions which allow us to reproduce both the Coulomb singularities in $r_i$ and in $r_{ij}$ and the correct asymptotic behavior of large distances.} As a result the wavefunction of the $^2 S_{1/2}$ Li ground state is written in the particular form
\begin{equation}
\label{ansatz}
 \psi = \mbox{\sf A} \left[ \phi(\vec{r}_1,\vec{r}_2,\vec{r}_3)\chi\right]\ ,
\end{equation}
with
\begin{equation}
\label{ansatzphi}
 \phi(\vec{r}_1,\vec{r}_2,\vec{r}_3) = f(r_1,r_2,r_3,r_{12},r_{13},r_{23})
 e^{- \alpha_1 r_1 - \alpha_2 r_2  -  \alpha_3 r_3 - \alpha_{12} r_{12}
- \alpha_{13} r_{13} - \alpha_{23} r_{23}} \,   ,
\end{equation}
where the pre-exponential factor  is a linear function of its
arguments, while $\alpha_i$ and $\alpha_{ij}$ are (non-linear) parameters.
${\sf{A}}$ is the three-particle {\it antisymmetrizer}
\begin{equation}
\label{Antiz}
{\sf A}=I - P_{12} - P_{13}- P_{23} + P_{231} + P_{312}\ .
 \end{equation}
Here $P_{ij}$ represents the permutation $i \leftrightarrow j$
and $P_{ijk}$ stands for the permutation of $123$ into $ijk$.
In Eq. (\ref{ansatz}), $\chi$ denotes a doublet spin eigenfunction ($S=1/2$) expressed as a linear combination
\begin{equation}
\label{ansatzspint}
\chi\ =\ \chi_1 + B \chi_2
 \end{equation}
of two linearly independent spin functions spanning the doublet spin space
of quantum numbers $S=1/2$, $M_s=1/2$:
\begin{equation}
\label{ansatzspin1}
  \chi_1 = 2^{-1/2}~ [\alpha(1) \beta(2) \alpha(3) - \beta(1) \alpha(2) \alpha(3)],
\end{equation}
and
\begin{equation}
\label{ansatzspin2}
  \chi_2 = 6^{-1/2} ~[ 2 \alpha(1) \alpha(2) \beta(3) - \beta(1) \alpha(2) \alpha(3) - \alpha(1) \beta(2) \alpha(3)] .
\end{equation}
In Eq.~(\ref{ansatzspint}), $B$ is a parameter which can be used to obtain the
optimum spin function, and $\alpha(i), \beta(i)$
are spin up,\,down functions of electron $i$.
In total, the function $\psi$ of Eq.~(\ref{ansatz}) is
characterized by seven parameters, plus any that may occur in the
pre-exponential factor $f$.

The matrix elements of $\cal H$ can be written as integrals over the
nine dimensions represented by $\vec{r}_1$, $\vec{r}_2$, $\vec{r}_3$.
Integrations over three dimensions describing overall orientation are
easily performed, and we end up with six-dimensional integrals over the
relative coordinates $(r_1,r_2,r_3,r_{12},r_{13},r_{23})$.  While it is
in principle possible to reduce these integrals analytically to one-dimensional integration: to expressions involving dilogarithm functions, as first shown by Fromm and Hill \cite{FrommHill}, the analytic properties of the resulting expressions were
found to be quite complicated (see Harris \cite{FEH}).  For that reason,
the primary method used in the present research was direct six-dimensional
numerical integration.

These numerical integrations were carried out using a suitable
partitioning of the ${\bf R^6}$ to subdomains based on a profile of
the integrand (for details, see e.g.\ Turbiner and Lopez \cite{turbiner:2006}). In each subdomain the numerical integration is done with a relative accuracy
of $\sim 10^{-5}$ to $10^{-6}$ by use of the adaptive D01FCF routine
from NAG-LIB \cite{NAGLIB} in a parallel manner. Due to the complicated
profiles of the integrands the numerical calculations are very difficult and if not
done with great care can  lead to a serious loss of accuracy.  By comparing
numerical and analytical evaluations of some of the simpler
matrix elements, it was verified that the numerical methods were
reliable at least to six significant digits.

Minimization of the energy with respect to the nonlinear parameters was
performed using the minimization package MINUIT from CERN-LIB \cite{CERNLIB}.

%\newpage

\section{Results}

The Li ground-state energies obtained for optimized variational wavefunctions
of the form given in Eqs.~(\ref{ansatz}) and (\ref{ansatzphi}) are
displayed in Table \ref{table1}; the corresponding
optimized variational parameters are given in Table \ref{table2}.
Each of the first seven rows of Table \ref{table1} describes a wavefunction
with a different pre-exponential factor; the last row of the table reports
the energy obtained from the most accurate existent Li ground-state
calculation \cite{exact}, a result extrapolated from a 9576-term
wavefunction of Hylleraas type and probably accurate to within $10^{-9}$ a.u.
Using these wave functions we calculated the variational energies and also the values of the cusp parameters:
\begin{equation}
\label{cusp}
  C\ =\ \frac{\langle\psi|\,\delta({\vec r})\frac{\partial}{\partial r}|
\psi\rangle}{\langle\psi|\,\delta({\vec r})|\psi\rangle}\ ,
\end{equation}
[cf. \cite{cusp}, Eq.~(18)], which for the exact wavefunction
should be equal to $-$3 when ${\vec r}={\vec r}_i$ (the electron-nuclear cusp).

\begin{table}[t]
\caption{Li ground-state energy $E$, cusp parameter $C_{eN}$
 [see Eq.~(\ref{cusp})] and the expectation values $<r_{ij}^{-1}>, <r_{ij}>$
 for the trial function in Eq.~(\ref{ansatz}) with various pre-factors $f$.}
\begin{displaymath}
\begin{array}{cccccl}
\hline\hline
 &  ~~~~~~~~f(r_1,r_2,r_3,r_{12},r_{13},r_{23})~~~~~~~~
 & ~\mbox{E (a.u.)} \  &\ -C_{eN}  &\ <r_{ij}^{-1}>  &\ <r_{ij}>   \\
\hline
\psi_1   &   1            & -7.4547 \ &\ 2.953   &\ 2.1732  &\ 10.0046    \\
\psi_2   & r_3                                           & -7.4712 \ &\ 2.958   &\ 2.1965  &\ 8.9553    \\
\psi_3   & r_{13}                                        & -7.4682 \ &\ 2.955   &\ 2.1922  &\ 8.9457   \\
\psi_4   & (1+ \beta_1 \ r_{3})\ \footnotemark[1]        & -7.4727 \ &\ 2.958   &\ 2.2091  &\ 8.6552   \\
\psi_5   & (1+ \gamma_1 \ r_{13})\ \footnotemark[2]      & -7.4686 \ &\ 2.955   &\ 2.1990  &\ 8.8953     \\
\psi_6^{*}   & { (1+ \beta_1 \ r_{3}+ \gamma_1 \ r_{13})}\
                                     \footnotemark[3] & { -7.4451} \  &\ { 2.901} &\ { 2.2501} &\ { 8.7688 } \\
\psi_6   & (1+ \beta_1 \ r_{3}+ \gamma_1 \ r_{13})\
                                     \footnotemark[4] & -7.4729 \  &\ 2.961 &\ 2.2027   &\ 8.7330\\
\hline
     \mbox{`Exact'} &                & -7.47806 \footnotemark[5] \ &\ \ 3.0\ &\ \ 2.1982 \footnotemark[6]\ &\ 8.6684 \footnotemark[6] \\
\hline\hline
\end{array}
\end{displaymath}
\hspace*{-80pt}
\begin{minipage}{4in}
\footnotetext[1]{\ $\beta_1 = -2.44486$}
\footnotetext[2]{\ $\gamma_1 =\ 9.53316$}
\footnotetext[3]{\ $\beta_1 = -3.82716,~ \gamma_1 = -0.47333$}
\footnotetext[4]{\ $\beta_1 = -2.77713,~ \gamma_1 = -0.26645$}
\footnotetext[5]{\ from Ref. \cite{exact} (rounded)}
\footnotetext[6]{\ from Ref. \cite{exact-exp} (rounded).}
\end{minipage}
\label{table1}
\end{table}

All the wavefunctions described in Table \ref{table2} display the expected
electronic shell structure; electrons 1 and 2 correspond to a
$1s^2$ pair, with an average exponential parameter that exhibits only a small
degree of screening relative to the bare-nucleus value $\alpha=3$.  Significant
energy improvement has been achieved by the ``split-shell'' description of
this electron pair (with $\alpha_1 > 3 > \alpha_2$).  Electron 3 (to zero order
the $2s$ electron) is not optimally described by a pure exponential, and
great improvement is obtained by giving it a Slater-type orbital (STO)
description, as in $\psi_2$, or even a hydrogenic $2s$ form, as in $\psi_4$.
Note that the sign of the pre-factor parameter $\beta_1$ produces the node
characteristic for a $2s$ orbital. The data for $\psi_3$ and $\psi_5$
show that inclusion of a linear interelectron distance improves
the variational energy (as indeed it must), but the improvement is not as
striking as that associated with the factor $r_3$.  Incidentally, most of
the improvement associated with the insertion of the factor $r_{13}$ in
$\psi_3$ simply reflects the fact that with high probability, $r_{13}$ is
similar in magnitude to $r_3$.  This observation becomes evident when one
notes that the $r_1$ distribution is $1s$-like and far more localized than is the
$2s$-like $r_3$ distribution.  Further flexibility, as
in $\psi_6$ (the best three-term prefactor) gives little additional gain over
that already achieved in $\psi_4$.  All the wavefunctions
also exhibit small negative values of the interelectron screening parameters
$\alpha_{ij}$, thereby improving the description of the repulsive
electron-electron correlation. {\it However, neglecting this screening by
 setting
all $\alpha_{ij}=0$ (see the function $\psi_6^{*}$, Table II) worsens
 significantly the variational energy as well as the cusp parameter,
 see Table I. Note that a function of this type was used in \cite{exact},
 yielding the most accurate variational energy thus far obtained,
  but at the expense of a very long expansion.}

\begin{table}[t]
\caption{Variational parameters $\alpha_i$ and $\alpha_{ij}$
 in [a.u.]$^{-1}$ and $B$ (dimensionless) for some trial functions from Table I.}
\begin{displaymath}
\begin{array}{cccccc}
\hline\hline
            & \hspace{9pt}\psi_1 & \hspace{9pt}\psi_2 & \hspace{9pt}\psi_4 & \hspace{9pt}\psi_6^{*} & \hspace{9pt}\psi_6  \\
\hline
\alpha_1     & ~~~~\hspace{9pt}3.2892~~~~ & ~~~~\hspace{9pt}3.3065~~~~ & ~~~~\hspace{9pt}3.3038~~~~ & ~~~~\hspace{9pt}3.3051~~~~
    & ~~~~\hspace{9pt}3.3044~~~~  \\
\alpha_2     & \hspace{9pt}2.3343 & \hspace{9pt}2.3291 & \hspace{9pt}2.3519 &  \hspace{9pt}2.0657 & \hspace{9pt}2.3603 \\
\alpha_3     & \hspace{9pt}0.4336 & \hspace{9pt}0.7004 & \hspace{9pt}0.7473 &
\hspace{9pt}0.6690 & \hspace{9pt}0.7327  \\
\alpha_{12}  & -0.2108 & -0.2050 & -0.2150 & 0. & -0.2218  \\
\alpha_{13}  & -0.0411 & -0.0311 & -0.0194 & 0. & -0.0227  \\
\alpha_{23}  & -0.0404 & -0.0316 & -0.0313 & 0. & -0.0276  \\
\hline
       B     & \hspace{13pt}0.06295 & \hspace{13pt}0.01416 & \hspace{7pt}-0.00201 & \hspace{7pt} 0.00277 & \hspace{7pt}-0.00202 \\
\hline\hline
\end{array}
\end{displaymath}
\label{table2}
\end{table}

All the results reported in this paper use the doublet spin function that
optimizes the trial energy for the given spatial function.
From the small values of $B$ in Table II,
we see that in every case the dominant contributor to the spin state
is (as expected) that which couples the $1s$ and $1s^\prime$ spatial
functions into a spin singlet.  However, inclusion of the other member
of the spin basis does influence the energy to some extent;  for
$\psi_1$ (the trial function with the largest optimum $B$), use of the second
spin state lowers the trial energy by about 0.001 a.u.

Recently, it was shown by one of the authors \cite{turbiner:AHO} that a
correct treatment of the domain of WKB asymptotics of the wavefunction
at large distances is very important for getting a high quality trial function. Usually, the large-distance asymptotic expansion of the
exponential phase of the wavefunction contains several terms that grow as the distance increases. All these terms should be reproduced in a trial function,
otherwise exponential deviation from the exact function at large
distances occurs. In the case of Lithium %%%%%%%[see (\ref{H})]
it can be verified that
\begin{equation}
\label{phase}
 \varphi\ \equiv \ - \log \psi\ =\ a_1 r_3 + a_2 \log r_3 +
 \mbox{ O} (1)\ ,\quad r_3 \rar \infty\ , \
   r_{1,2}\ \mbox{fixed} \ ,
\end{equation}
where $a_{1,2}$ are constants. These two terms in the expansion are the only
terms that grow as $r_3$ increases, and failure to reproduce them in the trial
function can lead to an exponentially large deviation of the trial function from the
exact eigenfunction at large (and intermediate) distances $r_3$. All
six of the functions we study reproduce at least the linear term in Eq.~(\ref{phase})
and all but $\psi_1$ reproduce both terms. This observation
might be considered as an
explanation why $\psi_1$ gives worse results for energy than the
other wavefunctions (see Table \ref{table1}).
Unfortunately, at this time a complete analysis of the
asymptotic behavior of the exponential phase at large distances
(in different directions in 6D $r$-space) is missing;  such an analysis
would be helpful for the identification of adequate trial functions.

Following the above analysis, it is not surprising that the function
$\psi_1$ gives the largest deviation from the exact values in both
the energy  $(\sim 0.5\%)$ and the cusp parameter
$C_{eN}$ $(\sim 1.6\%)$, while for the most accurate function $\psi_6$
these deviations are respectively $\sim 0.01\%$ and $\sim 1.3\%$.
It is worth noting that for the functions $\psi_{1-6}$ an increase of the accuracy in energy corresponds to a decrease of
the error in the cusp parameter $C_{eN}$ (see Table \ref{table1}).
A similar situation occurs for the expectation
value $\langle r_{ij} \rangle$
in comparison with the value reported in \cite{exact-exp}.
For $\langle r_{ij}^{-1}\rangle$ the largest deviation from
the result from \cite{exact-exp}
occurs for the function $\psi_1$ which also provides the largest deviation
for energy, cusp parameter and $<r_{ij}>$. We must emphasize
that all deviations in energy as well as in the
expectation values occur systematically at the third significant digit.

\section{Conclusion}

Simple and compact few-parameter trial functions are presented for the
ground state of the Lithium atom. These already provide a very accurate
ground state energy. These functions $\psi_{1-6}$ are the most accurate
among existent few-parameter trial functions.
However, the presented analysis does not seem final since one could explore
the more extended pre-factor
\[
  f(r_1,r_2,r_3,r_{12},r_{13},r_{23}) = (1 + \beta_1 \ r_1 + \beta_2 \ r_2
 + \beta_3 \ r_3 + \gamma_1 \ r_{23} + \gamma_2 \ r_{13}
 + \gamma_3 \ r_{12})  \ ,
\]
which seems beyond the computer resources presently available to us.

The wavefunctions used in the present work can be easily modified to study
excited states of the Lithium atom.

\begin{acknowledgments}

Two of the present authors (NLG and AVT) dedicate this work to
the third author (FEH) on the occasion of his 80th birthday.

NLG and AVT express their deep gratitude to J.C.~L\'opez Vieyra (ICN-UNAM)
for his interest in the work and for numerous useful discussions. NLG thanks
the Instituto de Ciencias Nucleares (UNAM, Mexico City) and AVT thanks
the Physics Department (UF, Gainesville) for the warm hospitality where
a part of the present work was done. AVT thanks
IHES (Bures-sur-Yvette, France) for the hospitality extended to him
where the present work was completed.
FEH acknowledges support of the U.S. National Science Foundation,
Grant PHY-0601758. The work of AVT was partially supported by
DGAPA {\bf IN121106-3} and CONACyT {\bf 47899-E, 58942-F} grants.
Actual computations were done on a 54-node FENOMEC cluster ABACO (IIMAS) at UNAM in M\'exico City.

\end{acknowledgments}

\end{document}